\documentclass[prb,twocolumn,showpacs,superscriptaddress,preprintnumbers,amsmath,amssymb]{revtex4}
\usepackage{graphicx}
\usepackage{dcolumn}
\usepackage{bm}
\usepackage{wasysym}

\newcommand{\be}{\begin{eqnarray}}
\newcommand{\ee}{\end{eqnarray}}

\def\beqn{\begin{eqnarray}}
\def\eeqn{\end{eqnarray}}
\def\beq{\begin{equation}}
\def\eeq{\end{equation}}
\begin{document}
\title{Geometric criticality between plaquette phases in integer-spin kagom\'e $XXZ$ antiferromagnets}
\author{Cenke~Xu}
\affiliation{Department of Physics, University of California,
Berkeley, CA 94720}
\author{J.~E.~Moore}
\affiliation{Department of Physics, University of California,
Berkeley, CA 94720} \affiliation{Materials Sciences Division,
Lawrence Berkeley National Laboratory, Berkeley, CA 94720}
\date{\today}
\begin{abstract}
The phase diagram of the uniaxially anisotropic $s=1$ antiferromagnet on the kagom\'e lattice includes a
critical line exactly described by the classical three-color model.  This line is distinct from the standard
geometric classical criticality that appears in the classical limit ($s \rightarrow \infty$) of the 2D $XY$ model;
the $s=1$ geometric $T=0$ critical line separates two unconventional plaquette-ordered phases that survive to
nonzero temperature.  The experimentally important correlations at finite temperature and the nature of the
transitions into these ordered phases are obtained using the mapping to the three-color model and a combination
of perturbation theory and a variational ansatz for the ordered phases.  The ordered phases show sixfold
symmetry breaking and are similar to phases proposed for the honeycomb lattice dimer model and $s=1/2$ $XXZ$
model.  The same mapping and phase transition can be realized also for integer spins $s \geq 2$ but then
require strong on-site anisotropy in the Hamiltonian.
\end{abstract}
\maketitle

\section{Introduction}
The geometric constraints that determine the infinite set of ground states in classical highly frustrated magnets are known in several celebrated instances to give rise to power-law or ``critical'' correlations at
vanishing temperature.  A frustrated magnet is one in which not all interaction energies can be simultaneously optimized; here we follow convention in describing by ``highly frustrated'' the class of frustrated magnets that have an infinite number of classical ground states, even after removing global symmetries.  An example of a frustrated magnet that is not highly frustrated is the triangular lattice XY antiferromagnet, which has a unique ground state once the three spins on a single triangle are fixed.

The classical XY antiferromagnet on the (2D) kagom\'e lattice and the classical Heisenberg antiferromagnet on the (3D) pyrochlore lattice are special members of the highly frustrated class: they show power-law correlations determined by a critical
two-component height model in the first case\cite{huse,henleyunpub,readunpub} and by 3D dipolar correlations in the second~\cite{youngblood}.  A general highly frustrated magnet, such as the Ising antiferromagnet on the kagom\'e or triangular lattice, has only short-ranged correlations, i.e., an exponential decay of correlations beyond a finite ``correlation length.''  Geometrically critical frustrated magnets can be thought of as on the boundary between disordered phases with short-ranged correlations and long-ranged ordered phases, but rather than being tuned by an external parameter, the criticality is a consequence of the fixed geometry of the lattice.  A major focus of current research in frustrated magnetism is how quantum effects in real magnets with finite spin $S$ modify these classical critical points, leading possibly to spin liquids~\cite{tchernyshyov,moessnerhex,nikolic,young}, emergent gauge symmetries~\cite{hermele}, and novel ordered phases.

The first goal of this paper is to show that the uniaxially anisotropic spin-1 antiferromagnet on the kagom\'e lattice has a range of parameters over which the ground states are exactly determined by the ``three-color model''~\cite{baxter}, which is the same geometrically critical model that appears in the {\it classical} XY model on the same lattice~\cite{huse,henleyunpub,readunpub}.  The $s=1$ kagom\'e antiferromagnet is thus a rare example where exact information can be obtained on a frustrated quantum magnet.  Uniaxial (XXZ) anisotropy is generic in kagom\'e materials because their layered quasi-2D structure implies XXZ anisotropy.  A typical kagom\'e compound is the $s=3/2$ material SrCr$_{8- }$Ga$_{4+x}$O$_{19}$~\cite{broholm}; nickel kagom\'e compounds have $s=1$.  A recent neutron scattering study~\cite{leekagome} of the iron jarosite KFe$_3$(OH)$_6$(SO$_4$)$_2$, which has $s=5/2$, observed, in addition to expected classical Heisenberg kagom\'e physics, several effects induced by Dzyaloshinkii-Moriya terms resulting from additional ions that break the inversion symmetry of the kagom\'e lattice.

The second part looks at the plaquette-ordered phases separated by this critical line, which serves as a
solvable starting point for perturbation theory.  The critical line occurs for reasons quite different from the
critical point of the classical XY model on this lattice; in particular, the $s=1$ problem has a nonvanishing
gap above the ground-state manifold in the thermodynamic limit, unlike the large-$S$ case.  Section II reviews the connection between the classical XY antiferromagnet and the three-color model and introduces the $XXZ$ model for
spin-1 antiferromagnets on the kagom\'e lattice.  From this we derive the existence of a critical line related to the three-color model.  Section III discusses the ordered phases separated by this critical line and the nature of the phase transition and concludes by developing a gauge theory description in order to compare this system to other recently studied models of frustrated magnetism.

\section{The kagom\'e antiferromagnet and the three-color model}

The ground states of the classical XY model on the kagom\'e lattice (Fig. 1) are
equivalent~\cite{huse,henleyunpub,readunpub}, up to an overall $O(2)$ rotation, to the states of the honeycomb
lattice ``three-color model''~\cite{baxter}.  Once one spin is fixed, all
other spins form angles of $0$ or $\pm {2 \pi /3}$ with this spin: labeling these three directions with colors, the ground states have one spin of each color around each small triangle (Fig. 1).  This is equivalent
to coloring the bonds of a honeycomb lattice with three colors in such a way that each vertex joins bonds of all
three colors.  
The ground-state degeneracy and chiral susceptibility were obtained by Baxter~\cite{baxter}.  Correlations within this set of (equally weighted) ground states, which determine the thermodynamics in the limit of low temperature, are determined by a two-component height model with an $SU(3)$ symmetry~\cite{huse,kondev}: one quantity with critical correlations is the chirality $\chi = \pm 1$ of colors around a vertex~\cite{huse,kondev,moorelee}: \beq \langle \chi(r_1) \chi(r_2) \rangle \sim {1 \over |r_1 - r_2|^4}.
\label{eqchiral} \eeq
Experimental observation of spin chirality in a kagom\'e compound with approximate Heisenberg symmetry~\cite{chalkershender} is discussed in~\cite{leekagome}.

Now consider $s=1$ spins on the sites of the kagom\`e lattice, with antiferromagnetic, uniaxially anisotropic interactions $J_{xy} \not = J_z$ between nearest neighbors.  Once the interactions are anisotropic,
symmetry requires the inclusion of on-site anisotropy at the same order: \beqn H &=& \sum_{\langle i j \rangle}
\left[ J_{xy} \left({S}^i_x \cdot {S}^j_x + {S}^i_y \cdot {S}^j_y \right) + J_z {S}^i_z \cdot {S}^j_z \right]
\cr && + D \sum_i \left({S}^i_z\right)^2. \label{anham}\eeqn Note that the on-site anisotropy term $D$ would be
constant in an $s=1/2$ $XXZ$ model.  Terms omitted for an isolated kagom\'e layer are long-ranged (beyond nearest-neighbor), involve more than 2 spin operators, or break time-reversal symmetry, so (\ref{anham}) is appropriate when exchange interactions are dominant and time-reversal is not explicitly broken.  Dzyaloshinskii-Moriya terms absent in (\ref{anham}) are allowed if the nonmagnetic lattice surrounding the kagom\'e layers breaks inversion symmetry, as in the compound studied in~\cite{leekagome}.

We start with the case $J_{xy}=0$, in which the model is effectively classical (i.e., $S_z$ on each site
commutes with the Hamiltonian).  Rewriting the Hamiltonian as a sum over the small triangles in the kagom\'e
lattice gives (here $1,2,3$ label the three spins in a triangle) \beqn H &=& \sum_\triangle \Big[ {J_z \over 2}
\left(S^1_z + S^2_z + S^3_z\right)^2 \cr && + {D - J_z \over 2} ((S^1_z)^2 + (S^2_z)^2 +(S^3_z)^2).
\Big]\label{zmodel} \eeqn The first term is minimized if $\sum_\triangle S^i_z = 0$, while the second term
favors $S_z=0$ for $D > J_z$ and $S_z = \pm 1$ for $D < J_z$.  This paper concentrates on the case of weak positive $D$, but first we briefly list the other possibilities.  For $D > J_z$, the ground state is simply $S_z = 0$ everywhere; the effective $U(1)$ gauge theory for low-energy excitations when $J_{xy} \not = 0$ is added to this case has been studied by Wen~\cite{wenkagome}.  For $D<0$, the ground state has $S_z = \pm 1$ everywhere and the problem reduces to the
classical Ising kagom\'e antiferromagnet, which is strongly disordered (i.e., the correlation functions in the equally weighted set of ground states fall off exponentially with distance).

The intermediate range, $0 < D < J_z$, is classically {\it critical}: a
ground state has on every triangle one spin with $S_z = 1$, one with $S_z = 0$, and one with $S_z = -1$, and this
condition is an exact statement of the three-color problem for the bonds of the honeycomb lattice.  Even though
the Hamiltonian $H$ {\it does not} have the additional $\mathbb{Z}_3$ color symmetry of the three-state Potts antiferromagnet or the classical XY antiferromagnet, the ground state family of states {\it does} have this
symmetry.  Note that the additional symmetry is $\mathbb{Z}_3$ rather than the permutation group of three colors $S_3$ because there is always a time-reversal symmetry in the Hamiltonian: the full permutation group of the three colors is made of this twofold symmetry that interchanges $S=+1$ and $S_z=-1$, plus three chirality-preserving rotations of the colors (isomorphic to $\mathbb{Z}_3$).  The icelike $T \rightarrow 0$ entropy per triangle of the kagom\'e lattice is half the entropy per hexagon of the three-coloring problem, $S_\triangle = 0.189557 k_B$~\cite{baxter}.

\begin{figure}
\includegraphics[width=2.5in]{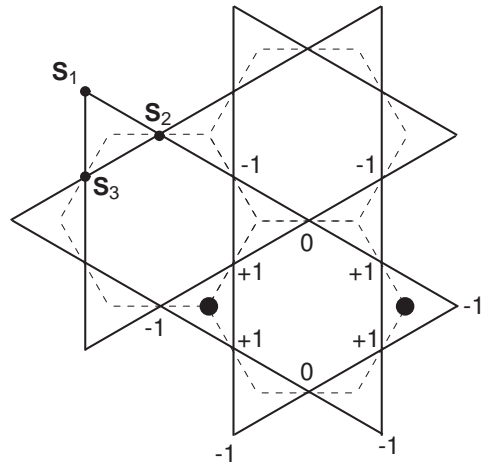}
\caption{A portion of the kagom\'e lattice, with the associated honeycomb lattice (dotted lines).  For
$0<D<J_z$, $J_{xy} = 0$, and $T = 0$, the three spins around every small triangle include one with $S_z = +1$,
one with $S_z = -1$, and one with $S_z = 0$.  The bottom right shows breakup of a single bond defect into two
vertex defects (large dots): the two defective vertices of the honeycomb lie on a loop containing only $S_z = 0,
+1$.} \label{figone}
\end{figure}

The key difference between this $s=1$ realization of the three-color model and the standard classical XY realization is the existence of an energy gap above the ground-state manifold in the $s=1$ model.  The existence of a critical line (a range of values of $D$ over which the model is critical) is simply a consequence of this gap: the set of ground states is exactly the same for all values of $D$, but the energy gap to defects varies with $D$ as now calculated.  Clearly the energy gap to these defects collapses at the two endpoints: below we obtain the correlation length at low temperatures induced by these defects, for which it is necessary to understand the nature of the long-ranged interaction between defects.

At nonzero temperature above this critical line, the free energy and correlation length can be determined using results on defects in the three-color model, although there are several signatures of the breaking of $\mathbb{Z}_3$ color
symmetry once states other than the ground states are considered.  Even though the lowest energy defect in a finite system consists of flipping an $S_z = 0$ site to
$S_z = \pm 1$, so the energy gap is $D$, the free energy correction at finite temperature does not go simply as
$\exp(-D/kT)$ because this local bond defect can break up into two vertex defects (Fig. 1).  For example, if we
flip $S_z = 0$ site to $S_z = 1$ site, then by switching $S_z = 1$ and $S_z = 0$ sites, this bond defect can
fractionalize into two $(1, 1, -1)$ vertex defects. This pair of vertex defects always lie in a loop made of
$S_z = 0$ and $S_z = 1$ sites, and are neither tightly bound nor completely free, but have a power-law entropic
interaction obtained by the Coulomb gas method~\cite{kondev} and verified in numerical transfer-matrix
studies~\cite{moorelee}.

For the model (\ref{zmodel}), the singular part of the free energy per site and correlation length are related
to the scaling dimension of lowest energy defects. The scaling dimension $x$ of the defect can be calculated
from the corresponding contour loop model, which gives $x=1/2$~\cite{kondev}. Following \cite{kdn,pt}, the
scaling of free energy can be calculated in terms of scaling dimension of defects,\beq f_s(T) \sim
|Y(m)Y(-m)|^{\frac{1}{2 - x}}.\eeq where $Y(m)$ and $Y(-m)$ are the fugacities of a pair of oppositely charged
defects.  Hyperscaling predicts that the singular part of the free energy per volume is determined by the inverse correlation volume: \beq f_s(T) \sim \xi(T)^{-2}. \eeq Hence the free energy and correlation length
behave at low temperature as \beq f_s(T) \sim e^{-{2 D / 3kT}},\quad \xi(T) \sim e^{D / 3kT}. \label{corr} \eeq
(These formulas describe the exponential part of the behavior: it is possible that there are power-law dependences in the prefactors.)  The chirality correlator (\ref{eqchiral}) will show power-law behavior for $r = |r_1-r_2| \ll \xi$, and then
decrease as $\exp(-r/\xi)$ once $r \gg \xi$. A small numerical transfer-matrix study (up to 9 bonds per unit cell) confirms qualitatively the prediction that $f_s$ for a small system of size $L \ll \xi(T)$ goes as $e^{-D/kT}$ when $kT \ll D$ with a temperature-dependent prefactor that increases with system size: in the thermodynamic limit, this prefactor gives rise to the different scaling in (\ref{corr}).

Recent interest in the classical three-color model has focused on its unusually slow dynamics at low
temperature, which invalidate local Monte Carlo methods~\cite{chakraborty,chamonunpub}: the $s=1$ anisotropic
kagom\'e antiferromagnet may provide an experimental system in which the dynamical phenomena predicted for this
classical model can be observed.  The rest of this paper concentrates on the quantum model: nonzero $J_{xy}$
generates two different quantum ordered phases separated by the classical line ($J_{xy} = 0$,$0<D<J_z$) with
critical correlations.  Unlike the power-law correlations on the critical line, the order parameters in these
phases are predicted to survive to $T>0$.  The following section starts with a heuristic derivation of the ordered phases, then
develop a lattice $U(1)$ gauge theory to justify some assumptions and connect to other models.

\section{Plaquette-ordered phases near the critical line}

\begin{figure}
\includegraphics[width=3.0in]{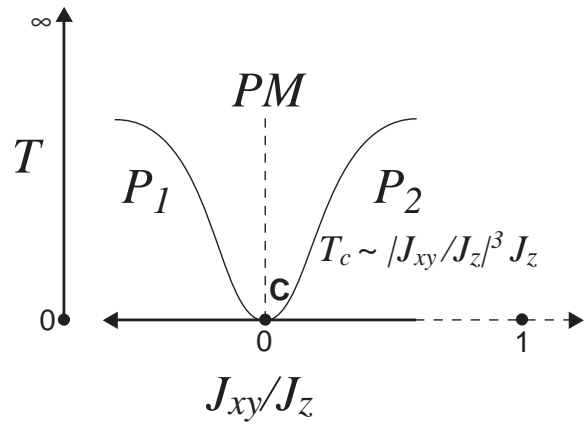}
\caption{Phase diagram of spin-1 $XXZ$ kagom\'e antiferromagnet for fixed $D \in (0,J_z)$.  The critical point
${\bf C}$ separates two distinct ordered phases $P_1$ and $P_2$ with sixfold symmetry breaking.  The correlation
length along the dotted line diverges as $T \rightarrow 0$ according to equation (\ref{corr}).} \label{figtwo}
\end{figure}

This section starts by finding states near the critical line using a standard perturbation-theoretic approach within the degenerate manifold of ground states found above.  The first splitting of the ground-state degeneracy of the three-color model occurs at third order in
perturbation theory in $J_{xy}/J_{z}$: the effective Hamiltonian is \beq H_{eff} = -t \sum_{\hexagon} (S^1_+
S^2_- S^3_+ S^4_- S^5_+ S^6_- + {\rm h.c.}) \label{effham}\eeq where $t = (3 J^3_{xy})/(2 J^2_z)$, and the
numbers $1$ to $6$ refer to consecutive spins around a hexagon of the kagom\'e lattice; spins live on bonds of the hexagons in Fig. 1. Similar loop terms occur in many models
when an effective $H$ is derived for states in the constrained space.  A similar perturbation theory has been carried out for the XXZ antiferromagnet on the triangular lattice~\cite{fazekas}.

$H_{eff}$ will induce resonances when a hexagon is ``flippable'', i.e., will superpose the two configurations
$S^i_z = (1,0,1,0,1,0)$ and $(0,1,0,1,0,1)$ around a hexagon, or superpose $(-1,0,-1,0,-1,0)$ and
$(0,-1,0,-1,0,-1)$.  Note that the third type of alternating hexagon, e.g. $(-1,1,-1,1,-1)$, is not flippable,
so that at this lowest nontrivial order of perturbation theory the $\mathbb{Z}_3$ symmetry of the classical
critical line is broken.  The existence of ${\it two}$ kinds of hexagons that can resonate will
lead to an additional symmetry breaking in the ordered phases we find, compared to predictions for the
$s=1/2$ $XXZ$ model~\cite{nikolicunpub} and the honeycomb quantum dimer model~\cite{moessnerhex,fradkin5}.

For nonzero $t$, we start by assuming that the system will favor a state with as many flippable hexagons as
possible.  This leads to the so-called $\sqrt{3}\times\sqrt{3}$ state, in which 2/3 of the hexagons are
flippable, as a classical starting point.  The hexagons in the honeycomb lattice form a triangular lattice with
three sublattices: in one $\sqrt{3}\times\sqrt{3}$ state, for example, sublattice $A$ has spins
(1,-1,1,-1,1,-1), sublattice $B$ has spins (1,0,1,0,1,0), and $C$ has (-1,0,-1,0,-1,0).  Now all hexagons of
sublattices $B$ and $C$ are flippable. To minimize the energy (\ref{effham}) starting from this state, one
expects resonance in the $B$ and $C$ sublattices, but both cannot simultaneously resonate without violating the
three-color constraint.  Consider a variational wavefunction in which only sublattice $B$ resonates.  The spins with
$S_z=-1$ are fixed and do not lie on the hexagons of sublattice $B$: call these sites ${\bar B}$.
The other spins resonate: the lowest-energy superposition is
\beqn \Psi_{var} &=& \prod_{i \in {\bar B}} |S^i_z = -1\rangle \cr &&
\otimes \prod_{{\hexagon} \in B} \left( { \left|\hexagon = {_0\ {^1}\ _0 \atop {^1}\ {_0}\ {^1}} \right\rangle
\pm \left|\hexagon = {_1\ {^0}\ _1 \atop {^0}\ _1\ {^0}} \right\rangle \over \sqrt{2}}  \right).
\label{wavefunc}\eeqn Here the sign in the superposition is the same for all hexagons, and for lowest energy has
the same sign as $t$ and $J_{xy}$: we label the state with $t<0$ as $P1$ and that with $t>0$ as $P2$.  This
state has $\langle H_{eff} \rangle = -|t|/3$ per hexagon of the whole lattice, since only 1/3 of hexagons
resonate.   It is an eigenstate of $H_{eff}$ if the Hilbert space is restricted to states satisfying the
three-color constraint.

These states correspond to sixfold symmetry breaking: hexagons on one of three sublattices, containing either $S_z = +1,0$ or $S_z = 0,-1$, resonate, and the other two sublattices are symmetric.  This combined sublattice and time-reversal symmetry breaking is expected to survive to finite temperature, with a transition temperature
$T_c$ of order $|t|$ (Fig. 2).  At finite temperature or by including additional terms in the model, it may be
possible to separate this sixfold symmetry breaking into two transitions.  The zero-temperature transition
between $P1$ and $P2$ is unusual in that it is a first-order transition by most definitions (the state changes
discontinuously at $t=0$), but with algebraic correlations at the transition point determined by the
$SU(3)$ theory of the three-color model: effectively the kagom\'e lattice is fine-tuned so that this scenario
occurs.  For integer spin $s \geq 2$, the three-color problem will again describe the $J_{xy}=0$ model if
sufficently strong ${S_z}^4$ anisotropy is added to the Hamiltonian.

A theory of two compact $U(1)$ gauge fields on the bonds of the honeycomb lattice gives insight into how
the ordered phases and critical line of this model connect to recent work on 2D and 3D
frustrated magnetism.  It is possible to use a single $U(1)$ gauge field, following the compact QED description
of the quantum dimer model \cite{fradkin90}: an integer-valued variable $E^i = S^i_z$ is assigned to each bond
$i$ of the honeycomb lattice to represent spin, with a gauge constraint that the sum of integers around a vertex
is 0, and an energetic constraint that restricts $E^i = -1,0,1$.  However, the theory now given in terms of two
gauge fields better captures the symmetries of the three-color line at $t=0$: the two gauge fields are related to the two components of the height-model description.

First let two {\it scalar} fields $E^i_a$, $a=1,2$, be assigned to each bond $i$: the values for the three spin
possibilities are \beq (E^i_1,E^i_2) = \begin{cases} (1,0)&{\rm if}\ S^i_z = 0\cr (-1/2,\sqrt{3}/2)&{\rm if}\
S^i_z = 1\cr (-1/2,-\sqrt{3}/2)&{\rm if}\ S^i_z = -1.
\end{cases}
\eeq These three possible states can be selected by allowing $(E_1,E_2)$ to range over the sites of a triangular
lattice centered on the origin with an additional energy ${1 \over k} \sum_i \left[(E^i_1)^2 +
(E^i_2)^2\right]$, $k \rightarrow 0$.  The Hilbert space is constrained to have the sum of $E$ fields vanish for the bonds around a vertex, creating two $U(1)$ gauge symmetries.

Next, note that a 2D unit vector ${\bf \hat n}_i$ can be assigned parallel or antiparallel to each bond $i$ so
that vertices of one sublattice of the honeycomb have 3 incoming bonds, while those of the other sublattice have
3 outgoing bonds. Now define two-component vector fields on bonds: ${\bf E}_a = E_a {\bf \hat n}_i$, $a = 1,2$.
Then the three-color constraint is Gauss's law with no background charges: $\nabla \cdot {\bf E}_a = 0$. Two
height fields $h_a$ on the faces are defined so that scalar $E_a$ on a bond is equal to $h^l_a - h^r_a$, where
$(r,l)$ indicate right and left w.r.t. the orientation on bonds: $E_a = ({\bf \hat z} \times {\bf \hat n}) \cdot
\nabla h_a$, which guarantees Gauss's law around each vertex. This definition of height variables corresponds to the standard definition in the classical 3-color model ~\cite{huse,kondev}.

The last step is to represent the ring-exchange terms, which break the $\mathbb{Z}_3$ symmetry.  Define conjugate operators on each bond $(A^i_1,A^i_2)$ with commutation relations \beq [A^j_a,E^k_b]
= i \delta_{jk} \delta_{ab}. \eeq Then $T^j_a \equiv \exp(i A^j_a)$ acts as a raising operator: it increases the
quantum number $E^j_a$ by 1.  This enables a compact representation of the ring-exchange terms proportional to
$t$: on bond $j$, $\exp(i A^j_a l^{(1)}_a)$ will raise $S^j_z = 0$ to $S^j_z = 1$ if $l^{(1)} =
(-3/2,\sqrt{3}/2)$.  Similarly, if $l^{(2)} = (-3/2,-\sqrt{3}/2)$ then $\exp(i A^j_a l^{(2)}_a)$ takes $S^j_z =
0$ to $S^j_z = -1$.  Defining vector fields ${\bf A}^j_a = A^j_a {\bf \hat n}_j$, the ring-exchange terms around
hexagons become \beq H^\prime = -2 t \left(\cos (\nabla \times {\bf A}_a l^{(1)}_a) + \cos(\nabla \times {\bf
A}_a l^{(2)}_a) \right). \label{ringex} \eeq Here, as usual in gauge theories of lattice spin models, the
meaning of $\cos(\nabla \times {\bf A})$ is that one takes the lattice circulation around a plaquette: for ${\bf
\hat v}_j$ an clockwise assignment of unit vectors along the bonds around a hexagon, \beqn \cos(\nabla \times
{\bf A}) &=& \cos(A_1 - A_2 + A_3 - A_4 + A_5 - A_6)\cr &=& \cos\left(\sum_{j=1,\ldots,6} {\bf \hat v}_j \cdot
{\bf A}_j \right). \eeqn Note that the two ring-exchange terms only act in one of two cases: either the spins
around the hexagon alternate between $S_z = 0$ and $S_z = 1$, or between $S_z = 0$ and $S_z = -1$.  The sixfold
symmetry breaking in the plaquette-ordered phases results from the absence of the possible ring-exchange term
with $l^{(3)} = (0,\sqrt{3})$.  Adding this term restores a $\mathbb{Z}_3$ color symmetry, with likely ninefold
symmetry breaking in the ordered phases.  The similarity of this gauge model to compact QED supports the conclusion that there is a gapped ordered state for $t \not = 0$.

A quantum height model can be obtained from this gauge theory: using the definition of height variables given above, the ring exchange term becomes $\sum_{i = 1}^2 -2 t \cos (\pi_a l^{(i)}_a)$, where $\pi_a$ is the operator conjugate to height. The quantum height Hamiltonian can be used to study a simple variational wave function that gives the same ordered state as (\ref{wavefunc}).  A generalization that can be studied using this technique is to an $s=1/2$ $XXZ$ kagom\'e bilayer in a magnetic field, with coupling between the layers that induces a three-color point: this system has columnar order rather than plaquette order.

\section{Conclusions}

We have used the proximity to a nontrivial classically solvable line to study part of the phase diagram of the
$s=1$ antiferromagnet on the kagom\'e lattice, including phases where $J_{xy} \not = 0$ and quantum Monte Carlo
studies are impractical.  The solvable line at zero temperature has an emergent $\mathbb{Z}_3$ symmetry of the three colors that does not exist in the full spectrum.  A promising way to extend this approach closer to the isotropic point ($J_{xy}=J_z,
D=0$) is via numerical series expansion in $J_{xy}/J_z$ from the critical line, to map out the extent of the plaquette-ordered phase.  Finding other cases where frustrated magnets have solvable critical points for finite $s$ similar to those in the kagom\'e and pyrochlore antiferromagnets at $s=\infty$ is an important direction for future research.

The unusual phase diagram of the spin-1 antiferromagnet on the kagom\'e lattice should be verifiable by experimental neutron scattering studies of such compounds: in particular, confirmation of the plaquette ordered state should be feasible if it survives for a significant range of values of $J_{xy}/J_z$.  The special low-temperature physics of the three-color model at $J_{xy}=0$ also can be discerned by sensitive thermodynamic measurements in principle, but away from the special three-color point, the zero-field thermodynamic signature of the phase transition into the plaquette-ordered phase will be similar to standard second-order phase transitions at finite temperature.

\acknowledgments

J. E. M. acknowledges helpful discussions and correspondence with C. Chamon, E. Fradkin, J. Kondev, and A.
Vishwanath.  The authors were supported by NSF DMR-0238760, the Hellman Foundation, and NERSC.

\bibliographystyle{../Newliou/apsrev}
\bibliography{../bigbib}

\begin{thebibliography}{24}
\expandafter\ifx\csname natexlab\endcsname\relax\def\natexlab#1{#1}\fi
\expandafter\ifx\csname bibnamefont\endcsname\relax
  \def\bibnamefont#1{#1}\fi
\expandafter\ifx\csname bibfnamefont\endcsname\relax
  \def\bibfnamefont#1{#1}\fi
\expandafter\ifx\csname citenamefont\endcsname\relax
  \def\citenamefont#1{#1}\fi
\expandafter\ifx\csname url\endcsname\relax
  \def\url#1{\texttt{#1}}\fi
\expandafter\ifx\csname urlprefix\endcsname\relax\def\urlprefix{URL }\fi
\providecommand{\bibinfo}[2]{#2}
\providecommand{\eprint}[2][]{\url{#2}}

\bibitem[{\citenamefont{Huse and Rutenberg}(1992)}]{huse}
\bibinfo{author}{\bibfnamefont{D.~A.} \bibnamefont{Huse}} \bibnamefont{and}
  \bibinfo{author}{\bibfnamefont{A.~D.} \bibnamefont{Rutenberg}},
  \bibinfo{journal}{Phys. Rev. B} \textbf{\bibinfo{volume}{45}},
  \bibinfo{pages}{7536} (\bibinfo{year}{1992}).

\bibitem[{\citenamefont{Henley}(1992)}]{henleyunpub}
\bibinfo{author}{\bibfnamefont{C.~L.} \bibnamefont{Henley}}
  (\bibinfo{year}{1992}), \eprint{unpublished}.

\bibitem[{\citenamefont{Read}()}]{readunpub}
\bibinfo{author}{\bibfnamefont{N.}~\bibnamefont{Read}}, \eprint{Kagom\'e
  workshop, unpublished (1992)}.

\bibitem[{\citenamefont{Youngblood and Axe}(1981)}]{youngblood}
\bibinfo{author}{\bibfnamefont{R.}~\bibnamefont{Youngblood}} \bibnamefont{and}
  \bibinfo{author}{\bibfnamefont{J.}~\bibnamefont{Axe}},
  \bibinfo{journal}{Phys. Rev. B} \textbf{\bibinfo{volume}{23}},
  \bibinfo{pages}{232} (\bibinfo{year}{1981}).

\bibitem[{\citenamefont{Tchernyshyov}(2004)}]{tchernyshyov}
\bibinfo{author}{\bibfnamefont{O.}~\bibnamefont{Tchernyshyov}},
  \bibinfo{journal}{J. Phys.: Cond. Mat.} \textbf{\bibinfo{volume}{16}},
  \bibinfo{pages}{S709} (\bibinfo{year}{2004}).

\bibitem[{\citenamefont{Moessner et~al.}(2001)\citenamefont{Moessner, Sondhi,
  and Chandra}}]{moessnerhex}
\bibinfo{author}{\bibfnamefont{R.}~\bibnamefont{Moessner}},
  \bibinfo{author}{\bibfnamefont{S.~L.} \bibnamefont{Sondhi}},
  \bibnamefont{and} \bibinfo{author}{\bibfnamefont{P.}~\bibnamefont{Chandra}},
  \bibinfo{journal}{Phys. Rev. B} \textbf{\bibinfo{volume}{64}},
  \bibinfo{pages}{144416} (\bibinfo{year}{2001}).

\bibitem[{\citenamefont{Nikolic and Senthil}(2005)}]{nikolic}
\bibinfo{author}{\bibfnamefont{P.}~\bibnamefont{Nikolic}} \bibnamefont{and}
  \bibinfo{author}{\bibfnamefont{T.}~\bibnamefont{Senthil}},
  \bibinfo{journal}{Phys. Rev. B} \textbf{\bibinfo{volume}{71}},
  \bibinfo{pages}{024401} (\bibinfo{year}{2005}).

\bibitem[{\citenamefont{Elstner and Young}(1994)}]{young}
\bibinfo{author}{\bibfnamefont{N.}~\bibnamefont{Elstner}} \bibnamefont{and}
  \bibinfo{author}{\bibfnamefont{A.~P.} \bibnamefont{Young}},
  \bibinfo{journal}{Phys. Rev. B} \textbf{\bibinfo{volume}{50}},
  \bibinfo{pages}{6871} (\bibinfo{year}{1994}).

\bibitem[{\citenamefont{Hermele et~al.}(2004)\citenamefont{Hermele, Fisher, and
  Balents}}]{hermele}
\bibinfo{author}{\bibfnamefont{M.}~\bibnamefont{Hermele}},
  \bibinfo{author}{\bibfnamefont{M.~P.~A.} \bibnamefont{Fisher}},
  \bibnamefont{and} \bibinfo{author}{\bibfnamefont{L.}~\bibnamefont{Balents}},
  \bibinfo{journal}{Phys. Rev. B} \textbf{\bibinfo{volume}{69}},
  \bibinfo{pages}{064404} (\bibinfo{year}{2004}).

\bibitem[{\citenamefont{Baxter}(1970)}]{baxter}
\bibinfo{author}{\bibfnamefont{R.~J.} \bibnamefont{Baxter}},
  \bibinfo{journal}{J. Math. Phys.} \textbf{\bibinfo{volume}{11}},
  \bibinfo{pages}{784} (\bibinfo{year}{1970}).

\bibitem[{\citenamefont{Broholm et~al.}(1990)\citenamefont{Broholm, Aeppli,
  Espinosa, and Cooper}}]{broholm}
\bibinfo{author}{\bibfnamefont{C.}~\bibnamefont{Broholm}},
  \bibinfo{author}{\bibfnamefont{G.}~\bibnamefont{Aeppli}},
  \bibinfo{author}{\bibfnamefont{G.~P.} \bibnamefont{Espinosa}},
  \bibnamefont{and} \bibinfo{author}{\bibfnamefont{A.~S.}
  \bibnamefont{Cooper}}, \bibinfo{journal}{Phys. Rev. Lett.}
  \textbf{\bibinfo{volume}{65}}, \bibinfo{pages}{3173} (\bibinfo{year}{1990}).

\bibitem[{\citenamefont{Grohol et~al.}(2005)\citenamefont{Grohol, Matan, Choi,
  Lee, Lynn, Nocera, and Lee}}]{leekagome}
\bibinfo{author}{\bibfnamefont{D.}~\bibnamefont{Grohol}},
  \bibinfo{author}{\bibfnamefont{K.}~\bibnamefont{Matan}},
  \bibinfo{author}{\bibfnamefont{J.-H.} \bibnamefont{Choi}},
  \bibinfo{author}{\bibfnamefont{S.-H.} \bibnamefont{Lee}},
  \bibinfo{author}{\bibfnamefont{J.~W.} \bibnamefont{Lynn}},
  \bibinfo{author}{\bibfnamefont{D.~G.} \bibnamefont{Nocera}},
  \bibnamefont{and} \bibinfo{author}{\bibfnamefont{Y.~S.} \bibnamefont{Lee}},
  \bibinfo{journal}{Nature Materials} \textbf{\bibinfo{volume}{4}},
  \bibinfo{pages}{323} (\bibinfo{year}{2005}).

\bibitem[{\citenamefont{Kondev and Henley}(1996)}]{kondev}
\bibinfo{author}{\bibfnamefont{J.}~\bibnamefont{Kondev}} \bibnamefont{and}
  \bibinfo{author}{\bibfnamefont{C.~L.} \bibnamefont{Henley}},
  \bibinfo{journal}{Nucl. Phys. B} \textbf{\bibinfo{volume}{464}},
  \bibinfo{pages}{540} (\bibinfo{year}{1996}).

\bibitem[{\citenamefont{Moore and Lee}(2004)}]{moorelee}
\bibinfo{author}{\bibfnamefont{J.~E.} \bibnamefont{Moore}} \bibnamefont{and}
  \bibinfo{author}{\bibfnamefont{D.-H.} \bibnamefont{Lee}},
  \bibinfo{journal}{Phys. Rev. B} \textbf{\bibinfo{volume}{69}},
  \bibinfo{pages}{104511} (\bibinfo{year}{2004}).

\bibitem[{\citenamefont{Chalker et~al.}(1992)\citenamefont{Chalker, Holdsworth,
  and Shender}}]{chalkershender}
\bibinfo{author}{\bibfnamefont{J.~T.} \bibnamefont{Chalker}},
  \bibinfo{author}{\bibfnamefont{P.~C.~W.} \bibnamefont{Holdsworth}},
  \bibnamefont{and} \bibinfo{author}{\bibfnamefont{E.~F.}
  \bibnamefont{Shender}}, \bibinfo{journal}{Phys. Rev. Lett.}
  \textbf{\bibinfo{volume}{68}}, \bibinfo{pages}{855} (\bibinfo{year}{1992}).

\bibitem[{\citenamefont{Wen}(2003)}]{wenkagome}
\bibinfo{author}{\bibfnamefont{X.-G.} \bibnamefont{Wen}},
  \bibinfo{journal}{Phys. Rev. B} \textbf{\bibinfo{volume}{68}},
  \bibinfo{pages}{115413} (\bibinfo{year}{2003}).

\bibitem[{\citenamefont{Kondev et~al.}(1996)\citenamefont{Kondev, de~Gier, and
  Nienhuis}}]{kdn}
\bibinfo{author}{\bibfnamefont{J.}~\bibnamefont{Kondev}},
  \bibinfo{author}{\bibfnamefont{J.}~\bibnamefont{de~Gier}}, \bibnamefont{and}
  \bibinfo{author}{\bibfnamefont{B.}~\bibnamefont{Nienhuis}},
  \bibinfo{journal}{J. Phys. A: Math. Gen.} \textbf{\bibinfo{volume}{29}},
  \bibinfo{pages}{6489} (\bibinfo{year}{1996}).

\bibitem[{\citenamefont{Nienhuis}(1987)}]{pt}
\bibinfo{author}{\bibfnamefont{B.}~\bibnamefont{Nienhuis}},
  \emph{\bibinfo{title}{Coulomb Gas Formulation of Two-dimensional Phase
  Transitions}} (\bibinfo{publisher}{Academic Press}, \bibinfo{year}{1987}).

\bibitem[{\citenamefont{Yin and Chakraborty}(2001)}]{chakraborty}
\bibinfo{author}{\bibfnamefont{H.}~\bibnamefont{Yin}} \bibnamefont{and}
  \bibinfo{author}{\bibfnamefont{B.}~\bibnamefont{Chakraborty}},
  \bibinfo{journal}{Phys. Rev. Lett.} \textbf{\bibinfo{volume}{86}},
  \bibinfo{pages}{2058} (\bibinfo{year}{2001}).

\bibitem[{\citenamefont{Castelnovo et~al.}(2004)\citenamefont{Castelnovo,
  Pujol, and Chamon}}]{chamonunpub}
\bibinfo{author}{\bibfnamefont{C.}~\bibnamefont{Castelnovo}},
  \bibinfo{author}{\bibfnamefont{P.}~\bibnamefont{Pujol}}, \bibnamefont{and}
  \bibinfo{author}{\bibfnamefont{C.}~\bibnamefont{Chamon}},
  \bibinfo{journal}{Phys. Rev. B} \textbf{\bibinfo{volume}{69}},
  \bibinfo{pages}{104529} (\bibinfo{year}{2004}).

\bibitem[{\citenamefont{Kleine et~al.}(1992)\citenamefont{Kleine, Fazekas, and
  Muller-Hartmann}}]{fazekas}
\bibinfo{author}{\bibfnamefont{B.}~\bibnamefont{Kleine}},
  \bibinfo{author}{\bibfnamefont{P.}~\bibnamefont{Fazekas}}, \bibnamefont{and}
  \bibinfo{author}{\bibfnamefont{E.}~\bibnamefont{Muller-Hartmann}},
  \bibinfo{journal}{Z. fur Physik B} \textbf{\bibinfo{volume}{86}},
  \bibinfo{pages}{405} (\bibinfo{year}{1992}).

\bibitem[{\citenamefont{Nikolic}()}]{nikolicunpub}
\bibinfo{author}{\bibfnamefont{P.}~\bibnamefont{Nikolic}},
  \eprint{cond-mat/0403332 (2004)}.

\bibitem[{\citenamefont{Fradkin et~al.}(2004)\citenamefont{Fradkin, Huse,
  Moessner, Oganesyan, and Sondhi}}]{fradkin5}
\bibinfo{author}{\bibfnamefont{E.}~\bibnamefont{Fradkin}},
  \bibinfo{author}{\bibfnamefont{D.~A.} \bibnamefont{Huse}},
  \bibinfo{author}{\bibfnamefont{R.}~\bibnamefont{Moessner}},
  \bibinfo{author}{\bibfnamefont{V.}~\bibnamefont{Oganesyan}},
  \bibnamefont{and} \bibinfo{author}{\bibfnamefont{S.~L.}
  \bibnamefont{Sondhi}}, \bibinfo{journal}{Phys. Rev. B}
  \textbf{\bibinfo{volume}{69}}, \bibinfo{pages}{224415}
  (\bibinfo{year}{2004}).

\bibitem[{\citenamefont{E.Fradkin and S.Kivelson}(1990)}]{fradkin90}
\bibinfo{author}{\bibnamefont{E.Fradkin}} \bibnamefont{and}
  \bibinfo{author}{\bibnamefont{S.Kivelson}}, \bibinfo{journal}{Mod. Phys.
  Lett} \textbf{\bibinfo{volume}{B4}}, \bibinfo{pages}{225}
  (\bibinfo{year}{1990}).

\end{thebibliography}
\end{document}